# Multiferroic $PbFe_{12}O_{19}$ Ceramics

*Guo-Long Tan*, Min Wang*

*State Key Laboratory of Advanced Technology for Materials Synthesis and Processing,*

*Wuhan University of Technology, Wuhan 430070, China*

**Abstract**

$PbFe_{12}O_{19}$ (PFO) powders in hexagonal structure have been synthesized by sol-gel process using lead acetate, glycerin and ferric acetylacetonate as the precursor. $PbFe_{12}O_{19}$ ceramics were obtained by sintering the $PbFe_{12}O_{19}$ powders at 1000℃ for 1 hour. Distorted flaky hexahedron grains are frequently observed in the SEM images of sintered $PbFe_{12}O_{19}$ ceramics. Large spontaneous polarization was observed in $PbFe_{12}O_{19}$ ceramic at room temperature, exhibiting a clear ferroelectric hysteresis loop. The remnant polarization of $PbFe_{12}O_{19}$ ceramic is estimated to be $Pr\sim21\mu C/cm^2$. The distortion of hexahedron grains as well as the Fe oxygen octahedron in its perovskite-like hexagonal unit cell is proposed to be the origin of polarization in $PbFe_{12}O_{19}$ ceramics. Meanwhile, $PbFe_{12}O_{19}$ ceramics demonstrate strong ferromagnetism at room temperature. Simultaneous occurrence of large ferroelectricity and strong ferromagnetism in $PbFe_{12}O_{19}$ ceramics holds promise for its application in new generation of electronic devices as a practical multiferroic candidate in single phase.

## 1  Introduction

In recent years, there has been increasing interest in multiferroic materials, which have simultaneous two or more order parameter in the same phase such as ferromagnetic, ferroelectric and/or ferroelastic ordering [1,2]. Because the couple between magnetic and electric ordering in multiferroic materials leads to electromagnetic effect, they provide wide range of potential applications such as multiple-state memory elements, novel memory media, transducer and new functional sensor [3].

It is clear now that the materials in which ferroelectricity and ferromagnetism coexist are rare [4,5] and mostly exhibit rather weak ferromagnetism. Because the room-temperature multiferroism is essential to the realization of multiferroic devices that exploit the coupling between ferroelectric and ferromagnetic orders at ambient conditions, $BiFeO_3$ together with more recently revealed $LuFe_2O_4$ [6-11] are currently considered to be promising candidates for practical device applications.  The perovskite $BiFeO_3$ is ferroelectric (Tc~1103K) and antiferromagnetism ($T_N$~643K), exhibiting weak magnetism at room temperature due to a residual moment from a canted spin structure [12], which could somehow prevent its practical application.  Therefore preparation of a material in which large ferroelectricity and strong ferromagnetism coexist would be a milestone for modern electrics and functionalized materials [13].  In the past, scientists were trying to look for multiferroic candidates in the traditional perovskite oxides, the typical example was $BiFeO_3$.  However, these perovskite oxides exhibit weak ferromagnetism.  In contrast, we start to explore the multiferroic candidates in an opposite direction, to look for ferroelectricity in the traditional ferromagnetic oxides, which holds perovskite-like lattice units.  $PbFe_{12}O_{19}$ is the one we are looking for.  As the electric polarization was found in materials of $YMnO_3$ [14], $ErMnO_3$ [15], $LiNbO_3$ and $BiFeO_3$ with hexagonal structure, it opens up a new direction for potential multiferroic candidate of such magnetic materials as $PbFe_{12}O_{19}$ having hexagonal structure.

M-type hexaferrites denoted as $PbFe_{12}O_{19}$ has attracted a lot of attention because of their excellent magnetic properties and potential application in various fields.[16,17] Their properties of large magneto crystalline anisotropy, high saturation magnetization and coercivity result in its usage as plastic magnets, recording media, permanent magnets, and components in microwave and high frequency devices.[18]  However, the ferroelectric aspect of $PbFe_{12}O_{19}$ has not been reported yet.  In this work, we are going to present the large spontaneous polarization of $PbFe_{12}O_{19}$ ceramic in addition to its strong ferromagnetism at room temperature.  The simultaneous occurrence of ferroelectricity and ferromagnetism in PFO ceramics will be discussed in detail in this paper.

## 2 Experimental procedure

Nanocrystalline $PbFe_{12}O_{19}$ powders were prepared by polymer precursor method using lead acetate ($Pb(CH_3COO)_2 \cdot 3H_2O$) and ferric acetylacetonate as starting materials. Typically, (0.7587g) lead acetate was dissolved in 15 ml glycerin to form a clear solution. After stirring at 60℃ for 1h, the solution was distilled in a rotary evaporator at 120℃ for 1h. In this way, the water in the solution was removed away. The distilled solution was stored in a three neck glass bottle. In order to prevent the hydrolysis of ferric acetylacetonate, the following experiments were carried out in the glove box. (6.3574g) ferric acetylacetonate was dissolved in 50ml anhydrous ethanol. Then the storing lead acetate solution was added into the ferric acetylacetonate solution at 50℃ under stirring for 1h. Here, the molar ratio of lead to iron ions was set to be less than 1:12 to balance the Pb loss. 10 mL ammonia and 15 mL polyethylene glycol mixture solution was added into the above solution, thus a colloid dispersion solution was formed. The dispersion solution was maintained at 353K under stirring for 8 hours. Afterwards, the colloid solution was moved out of the glove box. The water and organic solution were removed by centrifugation, the remaining colloid powders were calcined at 450℃ for 1.5h to completely remove the organic part. Then the powders were calcined again at 800℃ for 1h to yield pure PFO powder. 0.10g PFO powders were weighted and pressed in a module into pellet samples, which were then sintered at 1000℃ into PFO ceramics. Phase identification was performed by x-ray powder diffraction (XRD) method with Cu $K_\alpha$ radiation. The morphology and microstructure of pellet ceramics sintered at various temperatures were determined by SEM images, which are collected by a S-4800 field emission scanning electron microscopy (FESEM). The magnetization was measured using a Quantum Design physical property measurement system (PPMS). For P-E hysteresis loop measurement both surfaces of samples were coated with silver paste as electrodes, the ferroelectric hysteresis loop was measured using a home-made instrument, termed as ZT-IA ferroelectric measurement system.

# 3    Results and discussion

## *3.1. Structure and microstructure of PFO ceramics*

The XRD patterns of the PFO ceramic being sintered at 1000℃, together with its standard XRD pattern (PDF # 41-1373) are shown in Figure 1 (a) & (b).   The two XRD patterns match each other very well.   The XRD pattern showed the hexagonal structure of pure $PbFe_{12}O_{19}$ (PFO) ceramics, whose lattice parameter was determined to be a=5.8875Å & c=23.0639Å, respectively.   There is lattice contraction for the PFO ceramics in comparison with its standard lattice parameters ($a^*$=5.8945 Å & $c^*$=23.09 Å).   The magnitude of the lattice contraction was Δc= -0.026 Å & Δa= -0.007Å.   Obviously, the unit cell takes more contraction in c axis than that in a-axis.   The deviation of the lattice parameter away from its standard value indicates that there is a lattice distortion in the PFO ceramics being sintered at 1000℃.   This kind of lattice distortion could lead to the deformation of grains' shape and is beneficial to its spontaneous polarization, which creates ferroelectricity for the PFO ceramics.

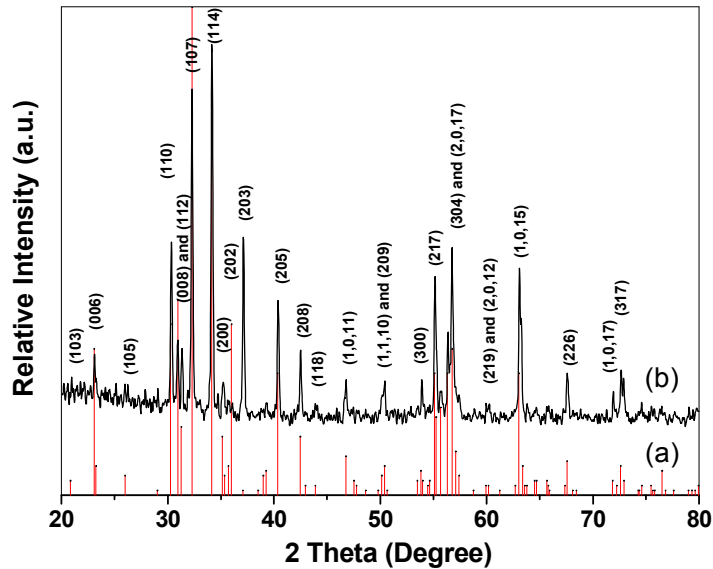

Figure 1: XRD patterns of the powders calcined at (a) 1000℃ for 1 hour and (b) standard pattern

*Figure 2* shows the field emission scanning electron microscopy (FESEM) images of the PFO ceramic being sintered at 1000℃.   It can be seen that the ceramic was well densified.   The grain size was estimated to be within the range of 0.5-3μm.   The distorted hexagonal flaky grains are frequently observed in the PFO ceramics, as being shown in *Figure 2* (b). Cylinder shaped grains can also be found in the SEM images. The hexahedron shaped grains are consistent with the hexagonal symmetry of PFO crystal. [0001] direction is normal to the surface of the flaky grains. The flaky shape suggests that the grains didn't take priority growth along with [0001] direction but with [0100] direction, which is the longest side of the flaky grains. Obviously the growth rate of the grains along with [0100] direction is much faster than that along with other directions. Therefore the grains did not take perfect hexahedron shapes, but actually take the deformed flaky ones. The deformation of the grains away from normal hexahedron reflects the distortion of hexagonal unit cells in PFO crystals. None centralized symmetry could exist in such distorted hexahedron grains. This kind of distortion is in agreement with the XRD measurement and may origin from the distortion of octahedron in the unit cell, which could be responsible for the ferroelectric behavior of PFO crystals.

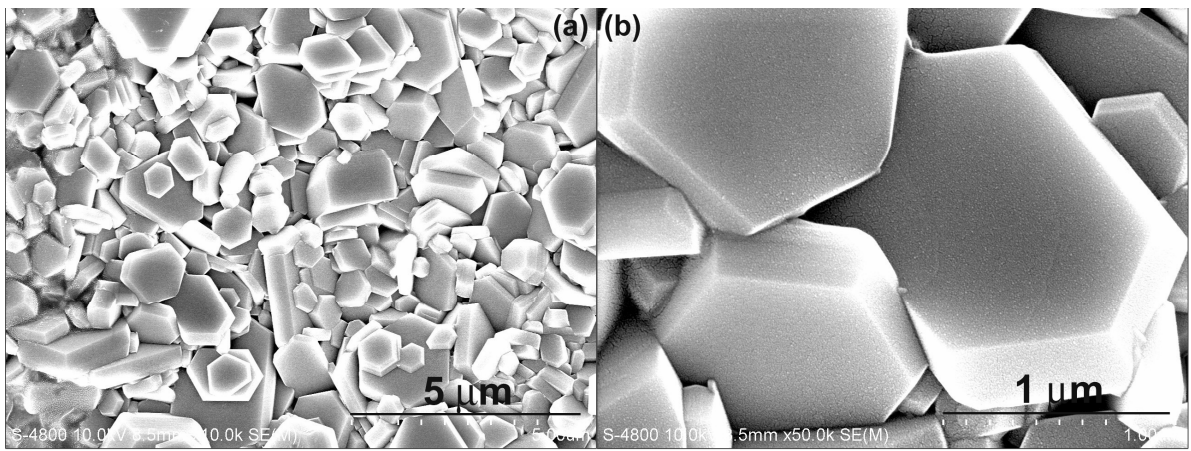

*Figure 2:* FESEM images of the ceramics sintered at 1000℃.

### 3.2. *Ferroelectric properties of PFO ceramics*

We investigated the effect of the constrained crystallographic state in the PFO ceramics sintered at 1000℃ on its physical property. Ferroelectric properties were characterized using polarization hysteresis and pulse polarization measurements. The specimen was parallel connected with a capacitor of 0.1μF for compensation when ferroelectric measurement was carried out at a frequency of 120 Hz for the PFO ceramics being sintered at 1000°C. Evidence for the characterization of ferroelectric state of PFO ceramic is provided in *Figure 3*, which shows a polarization cycle exhibiting a clear ferroelectric hysteresis loop. It can be seen from the electric hysteresis loop that the remnant polarization (Pr) is determined to be ~21μC/cm$^2$, which is around 3.5 times higher than the reported value of 6.1μC/cm$^2$ from BiFeO$_3$ ceramics [19]. To confirm this result, the polarization characteristic was measured under a pulsed trapezoidal probe condition, which is less likely to be convoluted by leakage and nonlinear dielectric effects. The pulsed remnant polarization (defined as $\Delta P = P^* - P^\wedge \approx 2P_r$, where $P^*$ is the switched polarization and $P^\wedge$ is the nonswitched polarization) shows a sharp increase of $\Delta P$ before 50 kilovolts, reaching a saturation value of about 63μC/cm$^2$ at 221 KV/m, as being shown in *Figure 4*. This polar state was found to be replicable, as evidenced by repeated experiments over several weeks.

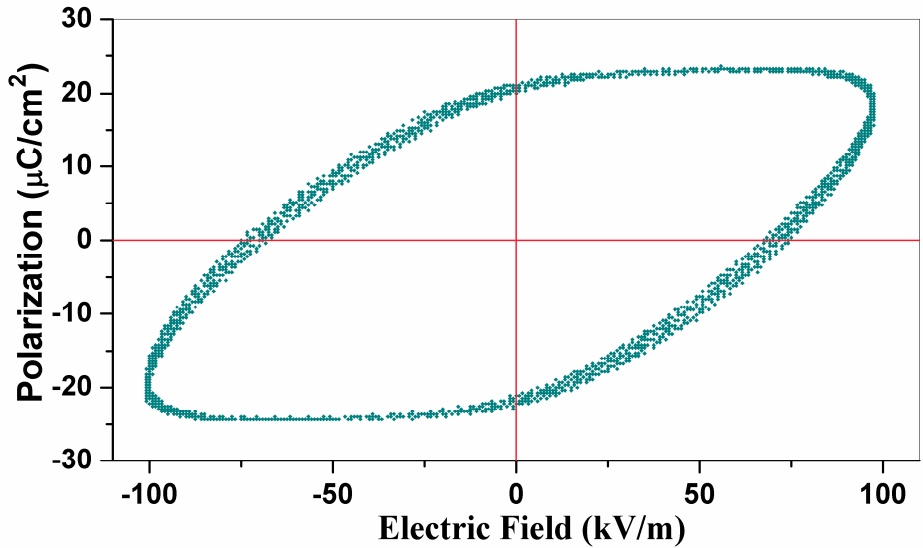

*Figure 3:* A ferroelectric hysteresis loop measured at a frequency of 120 Hz, which shows that the ceramics sintered at 1000℃ is ferroelectric with P$_r$~21μC/cm$^2$.

Another evidence to verify the ferroelectric characteristics of PFO ceramics is the polarization measurement at different frequencies. The value of the polarization at zero fields was not scaling with the frequency, which means that it does correspond to a remnant polarization rather than to a polarization due to leakage. Secondly, the polarization measurement at different frequencies within the range of 33~120 Hz and a fixed electric field (100V) demonstrates the reduction of remnant polarization with increasing frequency, which further confirms the ferroelectrics of PFO ceramics.

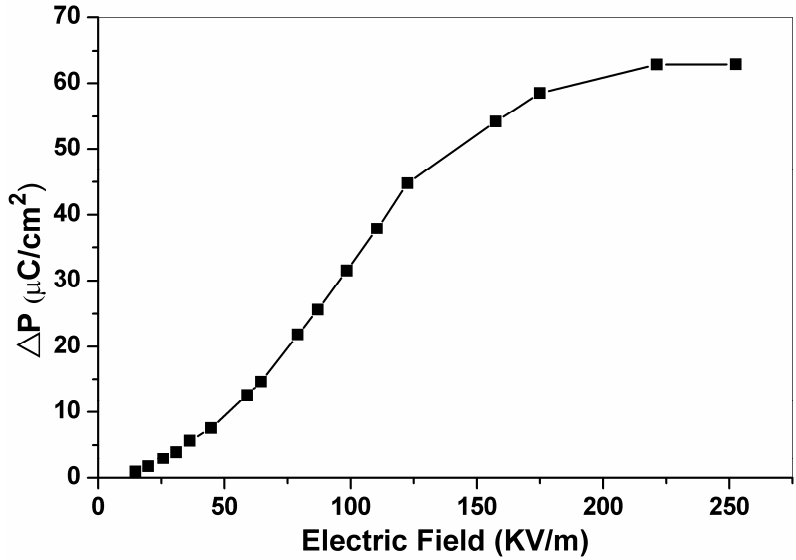

*Figure 4:* Polarization $\Delta P$ versus electric field measured with electrical pulses of 120 Hz.

The ferroelectrics usually exists in perovskite-structure oxides, such as $SrTiO_3$, $PbTiO_3$, which have the $BO_6$ oxygen octahedron. The ferroelectrics are also found in hexagonal compounds like $LiNbO_3$, which has two $NbO_6$ oxygen octahedrons in a unit cell. Below the Curie temperature, there is a distortion to a lower-symmetry phase accompanied by the shift off-center of the Nb cation and the shift off-public face of Li cation along the c-axis.

We firstly investigated the crystal structure model of $PbFe_{12}O_{19}$ (ICSD #36259) with space group P63/mmc, which is exhibited in *Figure 5*. Careful analysis of the unit model structure suggests a perovskite-like crystal structure with one distorted $FeO_6$ oxygen octahedron in hexagonal $PbFe_{12}O_{19}$ as being shown in *Figure 5*. Each hexagonal $PbFe_{12}O_{19}$ model has one $FeO_6$ oxygen octahedron in a sub-unit cell. In a normal octahedron, Fe cation is located at the center of an octahedron of oxygen anions. However, in the unit cell of $PbFe_{12}O_{19}$ below the Curie temperature, there is also a distortion to a lower-symmetry phase accompanied by the shift off-center of the small Fe cation. Fe cation shifts away from the center along b axis, while $O_3$ and $O_4$ shifts off their original positions of octahedron along opposite directions of a-axis, which leads to the distortion of $O_5$-Fe-$O_6$ bond away from straight line. The spontaneous polarization derives largely from the electric dipole moment created by the two shifts.

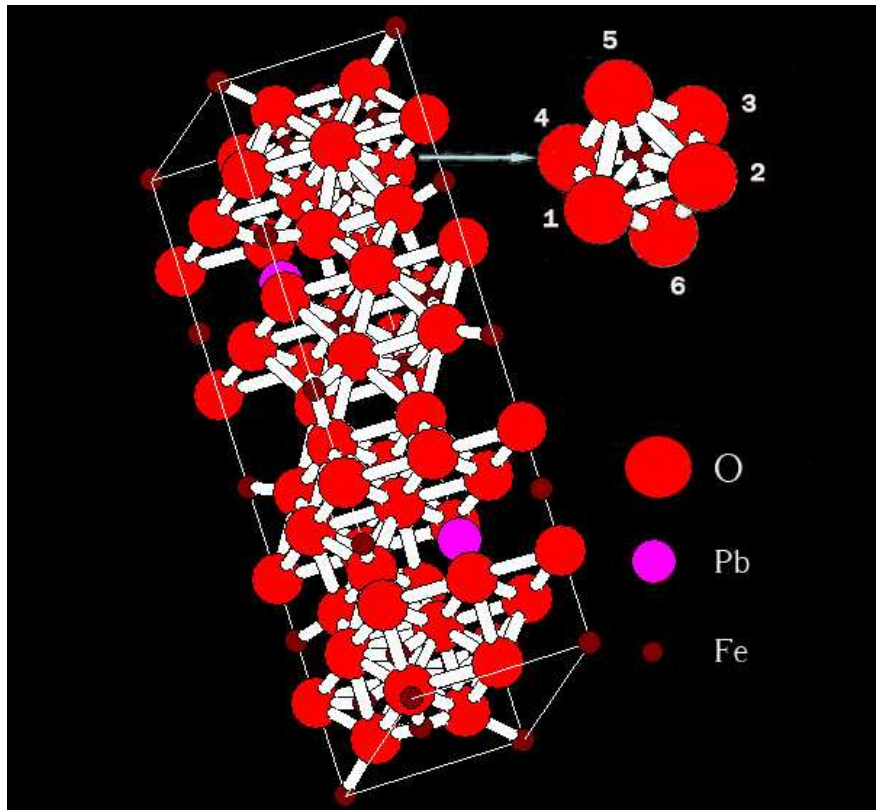

*Figure 5: Crystal structure model of PbFe$_{12}$O$_{19}$ (ICSD #36259) and magnified oxygen octahedron perovskite structure found in the crystal structure model of PbFe$_{12}$O$_{19}$.*

Meanwhile, distorted hexahedron grains are frequently observed in the sintered PFO ceramics, as being shown in SEM images of *Figure 2*. These grains are seriously deformed, deviating away from the centralized hexahedron. The non-central symmetrical octahedron in the hexagonal unit cell as well as the distorted hexahedron grains are proposed to be the origin of the electric dipole moments and thus be responsible for the spontaneous polarization of the PFO ceramic to the external electronic field, which behaves in a large ferroelectric hysteresis loop in *Figure 3*. The results demonstrate the influence of the distortion of structure on ferroelectric response in PFO ceramics.

### 3.3. *Ferromagnetism of PFO ceramics*

Now let's turn our attention to magnetic response. PbFe$_{12}$O$_{19}$ is a traditional ferromagnetic material, whose magnetization behavior has been widely studied. Magnetic properties of the PFO ceramics were measured at room temperature with a Quantum

Design physical property measurement system (PPMS).  The field-dependent magnetization hysteresis loop for PFO ceramic is shown in Figure 6.  The remnant magnetic ($M_r$) of the PFO ceramics was 24.5 emu/g, the saturation magnetization was ~50 emu/g.  The coercivity ($H_c$) of the PFO ceramics is 2168 Oe, which is lower than the reported value of 3800Oe of the PFO ceramics sintered at 800℃.[16]  The PFO ceramic exhibit reduced magnetization value in comparison with its counterpart being sintered at lower temperature.

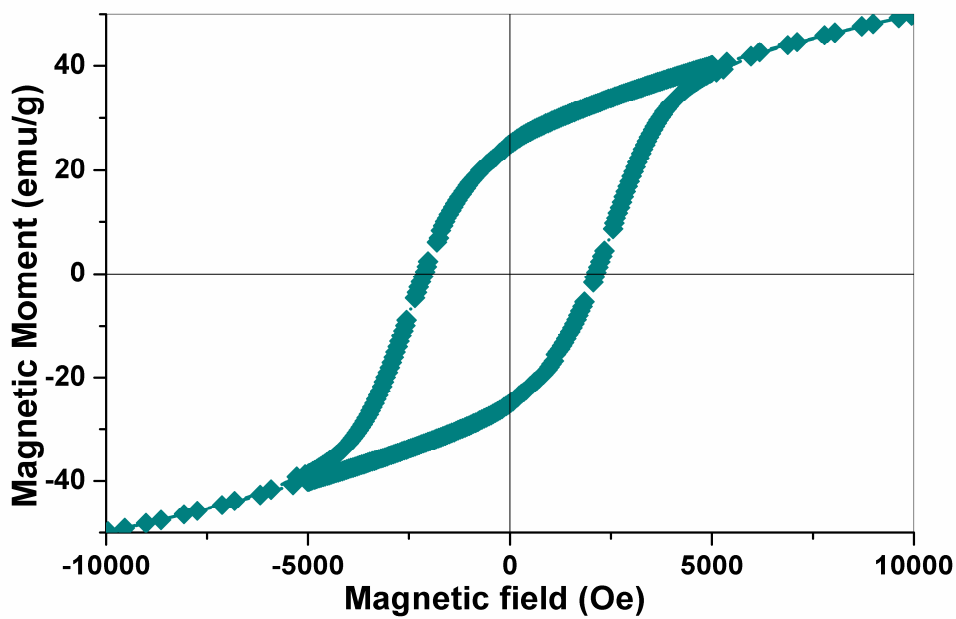

*Figure 6: Magnetic hysteresis loop of the $PbFe_{12}O_{19}$ ceramics sintered at 1000℃*

Size effect plays an important role on the reduction of the magnetization of PFO ceramics.  Usually the so-called critical size determines the properties of magnetic materials.[19-21]  When particle size is smaller than this critical value, particles locate within one single domain; otherwise multiple-domain may occur in particles.  As particles are larger than the single domain size for the ceramics being sintered at higher temperature, the domain walls become predominant.  The larger is particle size, the more are domain walls.  The domain walls could prevent somehow the switching of electron spins with external magnetic field, which results in reduction of magnetization of PFO ceramics when particle size increases with sintering temperature.

So far, $BiFeO_3$ is the best-known multiferroic material, which exhibits both ferromagnetic and ferroelectric ordering above room temperature. However, the weak ferromagnetism of BiFeO$_3$ may prevent its application. [22] In contrast, strong ferromagnetism and large ferroelectricity simultaneously occurred in our PFO ceramics above room temperature. This holds promise for its application in new generation of electronic devices as a practical multiferroic candidate in single phase.

# 4 Conclusion

In summary, pure lead hexaferrite (PbFe$_{12}$O$_{19}$) powders have been synthesized by a novel polymer precursor method using glycerin as solvent. The powders were pressed into pellets, which were sintered into ceramics at 1000℃ for 1 hour. The ferroelectric hysteresis loop of the sample shows that the remnant polarization (P$_r$) and the coercivity (H$_c$) are ~ 21 μC/cm$^2$ and 75 kV/m, respectively. The polarization $\Delta P$ versus electric field shows that the saturation polarization ($\Delta P$) of the sample could reach as high as ~63μC/cm$^2$. We propose that the source of polarization is the distortion of the Fe oxygen octahedron in the lattice unit of its perovskite-like hexagonal structure. The magnetic hysteresis loop of the sample shows that the remnant magnetic polarization (M$_r$) and the coercivity (H$_c$) are ~24.5 emu/g and 2168 Oe, respectively.